\documentclass[sigconf]{acmart}
\AtBeginDocument{%
  \providecommand\BibTeX{{%
    \normalfont B\kern-0.5em{\scshape i\kern-0.25em b}\kern-0.8em\TeX}}}

\copyrightyear{2022}
\acmYear{2022}
\setcopyright{rightsretained}
\acmConference[KDD '22]{Proceedings of the 28th ACM SIGKDD Conference on Knowledge Discovery and Data Mining}{August 14--18, 2022}{Washington, DC, USA}
\acmBooktitle{Proceedings of the 28th ACM SIGKDD Conference on Knowledge Discovery and Data Mining (KDD '22), August 14--18, 2022, Washington, DC, USA}
\acmDOI{10.1145/3534678.3539150}
\acmISBN{978-1-4503-9385-0/22/08}
\usepackage{amsmath,amsfonts,IEEEtrantools}

\usepackage{graphicx,caption,subcaption}

\usepackage{multirow}

\usepackage{microtype}

\begin{document}

\title[SoccerCPD: Formation and Role Change-Point Detection in Soccer Matches]{SoccerCPD: Formation and Role Change-Point Detection\\ in Soccer Matches Using Spatiotemporal Tracking Data}


\author{Hyunsung Kim}
\orcid{0000-0002-6286-5160}
\affiliation{%
  \institution{Fitogether Inc.}
  \city{Seoul}
  \country{South Korea}
}
\email{hyunsung.kim@fitogether.com}

\author{Bit Kim}
\affiliation{%
  \institution{Fitogether Inc.}
  \city{Seoul}
  \country{South Korea}
}
\email{bit.kim@fitogether.com}

\author{Dongwook Chung}
\affiliation{%
  \institution{Fitogether Inc.}
  \city{Seoul}
  \country{South Korea}
}
\email{dongwook.chung@fitogether.com}

\author{Jinsung Yoon}
\affiliation{%
  \institution{Fitogether Inc.}
  \city{Seoul}
  \country{South Korea}
}
\email{jinsung.yoon@fitogether.com}

\author{Sang-Ki Ko}
\orcid{0000-0002-5406-5104}
\affiliation{%
  \institution{Kangwon National University}
  \city{Chuncheon}
  \country{South Korea}
}
\email{sangkiko@kangwon.ac.kr}
\additionalaffiliation{%
  \institution{Fitogether Inc.}
  \city{Seoul}
  \country{South Korea}
}

\renewcommand{\shortauthors}{Hyunsung Kim et al.}

\begin{abstract}
  In fluid team sports such as soccer and basketball, analyzing team formation is one of the most intuitive ways to understand tactics from domain participants' point of view. However, existing approaches either assume that team formation is consistent throughout a match or assign formations frame-by-frame, which disagree with real situations. To tackle this issue, we propose a change-point detection framework named SoccerCPD that distinguishes tactically intended formation and role changes from temporary changes in soccer matches. We first assign roles to players frame-by-frame and perform two-step change-point detections: (1) formation change-point detection based on the sequence of role-adjacency matrices and (2) role change-point detection based on the sequence of role permutations. The evaluation of SoccerCPD using the ground truth annotated by domain experts shows that our method accurately detects the points of tactical changes and estimates the formation and role assignment per segment. Lastly, we introduce practical use-cases that domain participants can easily interpret and utilize.
\end{abstract}



\maketitle

\section{Introduction}
In fluid team sports such as soccer and basketball, players interact with each other while maintaining a role in a certain formation~\cite{Rein2016}. Since team formations highly affect the players' movement patterns, analyzing them is one of the most intuitive ways to understand tactics from the domain participants' point of view.

However, even with the recent acquisition of a vast amount of sports data, there remain several technical challenges for formation and role estimation as follows:
\begin{itemize}
    \item Coaches initially assign a unique role to each player, but they can change their instruction throughout the match.
    \item Players temporarily switch roles with their teammates.
    \item Abnormal situations such as set-pieces sometimes occur, in which all the players ignore the team formation.
\end{itemize}

Several studies~\cite{Bialkowski2014-Large, Narizuka2017, Narizuka2019, Shaw2019} have tried to estimate the team formation during a match using spatiotemporal tracking data, but they did not overcome all the above challenges. Bialkowski et al.~\cite{Bialkowski2014-Large} successfully captured the temporary role switches between players, but they have assumed that the team formation is consistent throughout half of a match. Narizuka et al.~\cite{Narizuka2017, Narizuka2019} found formations for each instant, resulting in too frequent formation changes that are not intended by coaches. Shaw et al.~\cite{Shaw2019} estimated a team formation per two-minute window, but their bottom-up approach has a trade-off between the reliability of detected change-points and the robustness against the abnormal situations. That is, since the endpoints of windows are the only candidate change-points, the discrepancy between predicted and actual change-points increases as the window size increases. On the other hand, a smaller window size becomes more sensitive to the situations such as set-pieces.

To fill this gap, we propose a change-point detection framework named \emph{SoccerCPD} that distinguishes tactically intended formation and role changes from temporary switches in soccer matches. First, we assign roles to players per frame by adopting the role representation proposed by Bialkowski et al.~\cite{Bialkowski2014-Large}. Next, we perform two-step change-point detections: (1) formation change-point detection (FormCPD) in each half of a match and (2) role change-point detection (RoleCPD) in each segment resulting from FormCPD.

In FormCPD, we formulate the players' spatial configuration---which we call the \emph{role topology} in this paper---at each frame as a role-adjacency matrix calculated from the Delaunay triangulation \citep{Narizuka2017}. Then we deploy a nonparametric change-point detection algorithm named discrete g-segmentation \citep{Song2020} to find change-points in this sequence of matrices. The detected change-points split a match into several \emph{formation periods} during which the team formation is assumed to be consistent. We represent the team's formation in each segment as the mean role-adjacency matrix and translate it into the language of domain participants (such as ``3-4-3'' or ``4-4-2'') by clustering all the obtained matrices.

In RoleCPD, we formulate the players' transposition by a sequence of role permutations in each formation period. As in the previous step, applying discrete g-segmentation to the permutation sequence with pairwise Hamming distances gives the role change-points. They split each formation period into several \emph{role periods} where the player-role assignment remains constant. Lastly, we give domain-specific ``position'' labels (such as ``left wing-back'' or ``center forward'') to the roles to enhance the interpretability.

With these results, we make the following four contributions:
\begin{itemize}
    \item applying a state-of-the-art change-point detection method to the sports tracking data, which discovers the underlying tactics and their changes in an unsupervised manner.
    \item effectively representing the spatial configuration and role switches of players in a time interval as a mean role-adjacency matrix and a sequence of permutations, respectively.
    \item suggesting practical applications such as switching pattern mining and automatic set-piece detection, which can be easily interpreted and utilized by domain participants.
    \item providing a Python implementation of our framework with sample GPS data\footnote{https://github.com/pientist/soccercpd}.
\end{itemize}

\section{Related Work}
\subsection{Change-Point Detection}
Change-point detection (CPD) is the problem of finding points when the governing parameters of the process change. CPD methods are classified into parametric and nonparametric ones depending on whether they specifically estimate the parameters or directly find change-points \cite{Truong2020}.

Parametric CPD models assume a certain form of distribution for given observations and find change-points of the parameters. They include maximum likelihood estimations~\cite{Ko2015, Pein2017, Lavielle2000}, piecewise linear models~\cite{Bai1994, Bai1997, Qu2007}, Mahalanobis metric-based models~\cite{Lajugie2014}, and so on. However, they have limitations in that they are sensitive to the violation of the assumptions of distributions \cite{Aminikhanghahi2017}.

Nonparametric CPD methods directly detect breakpoints without estimating distribution parameters. In general, they are based on nonparametric statistics such as rank statistics~\cite{Lung-Yut-Fong2015}, graph-based scan statistics~\cite{Chen2015, Chu2019, Song2020}, and kernel estimation~\cite{Shawe-Taylor2004, Harchaoui2009, Gretton2012}. Free from the above assumptions of distributions, they have shown greater success for various datasets~\cite{Aminikhanghahi2017}. Particularly, graph-based methods~\cite{Chen2015, Chu2019, Song2020} are considered more flexible in real applications since they only use similarity information and thus are applicable to high-dimensional data or even non-Euclidean data.

\subsection{Formation Estimation in Team Sports}
Several studies have analyzed spatiotemporal tracking data to estimate team formations and find the roles of players in team sports. Bialkowski et al.~\cite{Bialkowski2014-Large} proposed the ``role representation'' that dynamically assigns a unique role to each player frame-by-frame. Narizuka et al.~\cite{Narizuka2017} expressed a temporary formation as an undirected graph from Delaunay triangulation~\cite{Delaunay1934} and clustered the graphs obtained from a match to characterize the formation. Later, Narizuka et al.~\cite{Narizuka2019} combined their method with the role representation to cluster formations over multiple matches. Also, Shaw et al.~\cite{Shaw2019} used relative positions between players to estimate team formation per two-minute window.

Moreover, there have been other approaches relying on the annotated or estimated formation for deeper analyses. Bialkowski et al.~\cite{Bialkowski2014-Win} and Tamura et al.~\cite{Tamura2015} studied the correlation between team formation and other context information such as the existence of home advantage or the result of the previous match. Machado et al.~\cite{Machado2017} and Wu et al.~\cite{Wu2019} suggested visual interfaces showing the dynamic changes of the team formation. Especially, many studies have utilized the result of role representation above for various purposes such as highlight detection~\cite{Wei2013}, similar play retrieval~\cite{Sha2016}, team style identification~\cite{Bialkowski2014-Identifying}, and player style identification~\cite{Kim2021}.

\section{Problem Definition} \label{problem_def}
Soccer coaches change the team formation up to several times during a match, or they command role switches between players without formation change. Thus, our method is divided into two steps to track their decisions: first finds formation change-points and then finds role change-points. In this section, we formalize each step as a separate CPD problem.

\begin{figure*}
  \includegraphics[width=\textwidth]{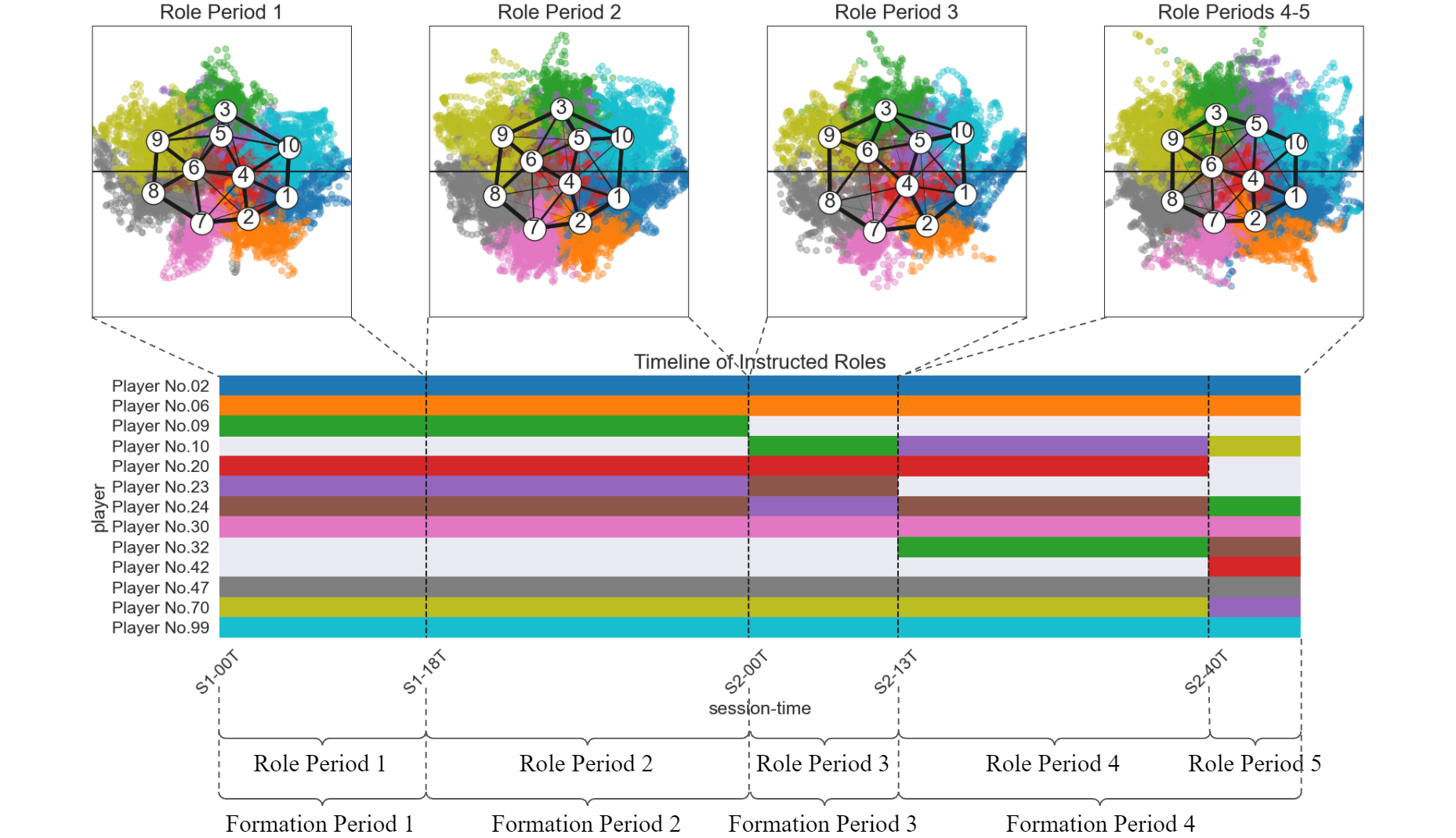}
  \caption{An example of formation and role change timeline as a result of SoccerCPD.}
  \Description{The upper figures indicate the team formation per formation period. The points of each color indicate the frame-by-frame coordinates of a certain role normalized by the team's mean location. The white circles are the mean locations of individual roles and are connected by edges whose widths are the corresponding elements in the mean role-adjacency matrices. The timeline in the lower part indicates the roles assigned to players per role period.}
  \label{fig:timeline}
\end{figure*}

\subsection{Formation Change-Point Detection} \label{formcpd_def}
One can consider team formation as a probability distribution over several role topologies. Given the sequence of observations in the form of players' $(x,y)$ locations in 10 Hz, the goal of FormCPD is to find change-points of this distribution. Thus, we first need to find effective features representing role topology from raw observations and then perform CPD using the sequence of the features.

Formally speaking, given a sequence of features $\{ A(t) \}_{t \in T}$ indicating the role topology, we aim to find a partition $T_1 < \cdots < T_m$ of $T$ such that $t \in T_i$ implies $A(t) \sim \mathcal{F}_i$ for some distinct distributions $\mathcal{F}_1, \ldots, \mathcal{F}_m$. Here we call each interval $T_i$ a \emph{formation period} during which the team maintain a certain formation represented as the distribution $\mathcal{F}_i$.

\subsection{Role Change-Point Detection} \label{rolecpd_def}
Bialkowski et al. \cite{Bialkowski2014-Large} proposed the role representation that gives the frame-by-frame role assignment to the outfield players, but they did not distinguish long-term tactical changes from these temporary role swaps. Given that a player generally possesses a constant role instructed by the coach for a certain period, the goal of RoleCPD is to find change-points where long-term tactical changes occur. That is, we further partition a formation period $T_i$ into several time intervals $T_{i,1} < \cdots < T_{i,n_i}$ named \emph{role periods} where each player possesses a constant role during a single role period. Then, we can deem the frame-by-frame role interchanges (the result of the role representation) inside role periods as temporary swaps.

Formally speaking, for the set $P$ of players in a match time $T$, the role representation finds a set $\mathcal{X} = \{ X_1, \ldots , X_N \}$ of roles and \emph{player-to-temporary-role} (P-TR) maps $\{ \beta_t : P \rightarrow \mathcal{X} \}_{t \in T}$ such that
\begin{equation}
    \beta_t(p) \neq \beta_t(q) \quad \forall p \neq q \in P,\ t \in T
\end{equation}
Here we aim to express the given P-TR maps $\{ \beta_t \}_{t \in T}$ as the composition of \emph{player-to-instructed-role} (P-IR) maps $\{ \alpha_t : P \rightarrow \mathcal{X} \}_{t \in T}$ and \emph{temporary role permutations} (RolePerm) $\{ \sigma_t : \mathcal{X} \rightarrow \mathcal{X} \}_{t \in T}$ with $\sigma_t \in \mathrm{S}(\mathcal{X})$ (symmetric group on $\mathcal{X}$), i.e., $\beta_t = \sigma_t \circ \alpha_t$. In other words, given P-TR maps $\{ \beta_t \}_{t \in T}$ in $T_i$, we find a partition $T_{i,1} < \cdots < T_{i,n_i}$ of $T_i$ and P-IR maps $\{ \alpha_t \}_{t \in T}$ satisfying the followings:

\begin{itemize}

    \item (\emph{Period-wise consistency}) The instructed role of every player is constant during each role period $T_{i,j}$. i.e.,
    \begin{equation}
        \alpha_t(p) = X_p^{(i,j)} \quad \forall t \in T_{i,j}
    \end{equation}
    for some $X_p^{(i,j)} \in \mathcal{X}$.
    
    \item (\emph{Uniqueness}) No two players are instructed to take the same role in a role period. i.e.,
    \begin{equation}
        X_p^{(i,j)} \neq X_q^{(i,j)} \quad \forall p \neq q \in P
    \end{equation}
    
    \item (\emph{Existence of a role change}) A change of role period implies a change of instruction. i.e., for each $j \in \{ 1, \ldots , n_i-1 \}$, there exists $p \in P$ such that
    \begin{equation}
        X_p^{(i,j)} \neq X_p^{(i,j+1)}
    \end{equation}
    
\end{itemize}
Note that the last condition is valid only for the change \emph{inside} a single formation period. Namely, adjoining role periods from distinct formation periods can have the same player-role assignment.

\section{Formation Change-Point Detection} \label{formcpd}
At first, we express the role topology at each frame as a role-adjacency matrix calculated from the Delaunay triangulation (Section~\ref{formcpd_seq}). Then, we detect formation change-points by applying discrete g-segmentation on the sequence of role-adjacency matrices (Section~\ref{formcpd_gseg}). Finally, we represent the formation in each segment as the mean role-adjacency matrix and cluster the matrices from the entire dataset to give domain-specific labels such as ``3-4-3'' or ``4-4-2'' to individual formation periods (Section~\ref{formcpd_cluster}). Please refer to the block diagram in Appendix~\ref{pipeline} for better understanding of the whole pipeline.

\subsection{Calculating the Sequence of Role-Adjacency Matrices} \label{formcpd_seq}
First, we assign roles to the players per frame by the role representation proposed by Bialkowski et al.~\cite{Bialkowski2014-Large}. It groups the player locations recorded throughout a session into role clusters, under the constraint that no two players take the same role at the same time. This is achieved by the EM algorithm below:
\begin{itemize}
    \item Initialization: Assign a unique role label from 1 to $N$ to each outfield player and estimate the density of each role.
    \item E-step: Apply the Hungarian algorithm~\cite{Kuhn1955} per frame to reassign roles to players using the cost matrix based on the log probability of a role generating a player location.
    \item M-step: Update the density parameters based on the role labels reassigned in E-step.
\end{itemize}

\begin{figure}[htb]
    \centering
    \begin{subfigure}[t]{0.23\textwidth}
    	\includegraphics[width=\textwidth]{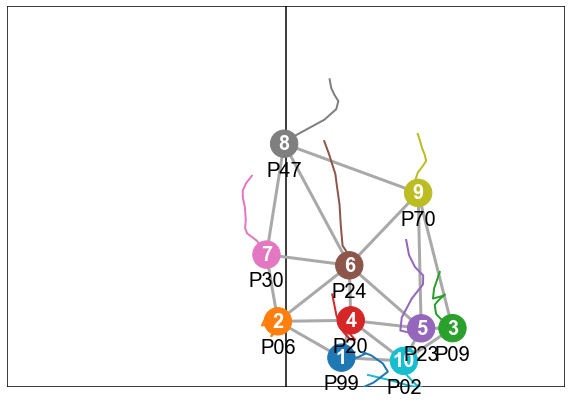}
    	\caption{} \label{fig:temp_mat}
    \end{subfigure}
    \begin{subfigure}[t]{0.23\textwidth}
    	\includegraphics[width=\textwidth]{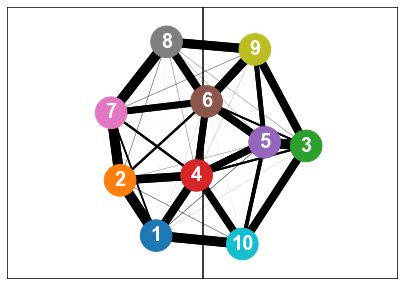}
    	\caption{} \label{fig:mean_mat}
    \end{subfigure}
    \caption{(a) A temporary Delaunay graph with player trajectories at a certain frame and (b) the weighted graph drawn from the mean role locations and the mean role-adjacency matrix for a formation period. The number inside each circle indicates a role label from 1 to 10, while the number below each circle indicates the player's uniform number.}
\label{fig:adj_mats}
\end{figure}

Next, we perform Delaunay triangulation \citep{Delaunay1934} to get the adjacency matrices $\{A(t)\}_{t \in T} \subset \mathbb{R}^{N \times N}$ between the role labels. That is, the components $a_{kl}(t)$ are defined as
\begin{equation*}
    a_{kl}(t) = 
    \begin{cases}
        1 & \text{if the \emph{roles} $X_k$ and $X_l$ are adjacent at $t$},\\
        0 & \text{otherwise}
    \end{cases}
\end{equation*}
for the set $\mathcal{X} = \{ X_1, \ldots , X_N \}$ of roles. Fig.~\ref{fig:temp_mat} shows an example of Delaunay graph on role locations.

The idea of applying Delaunay triangulation to player tracking data was first proposed by Narizuka et al. \cite{Narizuka2017}. However, they used uniform numbers as indices of the adjacency matrices and applied the role representation only for aligning uniform numbers from multiple matches to a single set of indices \citep{Narizuka2019}. While this method can discover position-exchange patterns well, the resulting player-adjacency matrices are highly affected by temporary switches or irregular situations such as set-pieces which are irrelevant to the original team formation. Hence, we use role labels instead of player labels as the indices of adjacency matrices, which are a more robust representation of the team's spatial configuration.

\subsection{Applying Discrete g-Segmentation to the Sequence of Matrices} \label{formcpd_gseg}

In order to find formation change-points, we separately apply CPD to the sequence of role-adjacency matrices obtained from each half of a match.

Since our sequences are high-dimensional ($N \times N$ matrices for FormCPD with $N = 10$ in general) or even non-Euclidean (permutations for RoleCPD in Section~\ref{rolecpd}), we focus on graph-based CPD~\cite{Chen2015, Chu2019, Song2020}. They draw a similarity graph such as the minimum spanning tree on the observations and calculate a scan statistic $R(t)$ for each time $t$ by checking how many edges connect observations before and after $t$. Since a large difference between the observations before and after $t$ leads to large $R(t)$, one can regard $\tau = \arg\max_t R(t)$ as a possible change-point if $R(\tau)$ exceeds a certain threshold after normalization. As they only use similarity information between observations, they are applicable to our sequences with proper distance functions.

Another issue is that our sequences have repeated observations (binary matrices for FormCPD here and permutations for RoleCPD in Section~\ref{rolecpd}), which makes the optimal similarity graph not uniquely defined. Among the series of graph-based CPD methods, discrete g-segmentation~\cite{Song2020} can only cover this case by using the average statistic $R_{(a)}(t)$ or the union statistic $R_{(u)}(t)$ introduced by Zhang and Chen~\cite{Zhang2017} as alternatives for $R(t)$. That is, they obtain a unique scan statistic either by ``averaging'' the scan statistics over all optimal graphs or by calculating the scan statistic on the “union” of all optimal graphs. Therefore, we employ the discrete g-segmentation to find change-points in our discrete high-dimensional/non-Euclidean sequences. (Please refer to Appendix~\ref{baseline_comparison} that empirically justifies the choice of CPD method.)

To elaborate, we use the $L_{1,1}$ matrix norm
\begin{equation}
    d_M(A(t), A(t')) = \| A(t) - A(t')\|_{1,1}
    = \sum_{k=1}^N \sum_{l=1}^N |A_{kl}(t) - A_{kl}(t')|
    \label{eq:manhattan}
\end{equation}
(which we call \emph{Manhattan distance} hereafter) as the distance measure between role-adjacency matrices. Also, we compute the \emph{switch rate} at $t$ by counting the number of players whose temporary roles differ from their most frequent roles in the session. Then, we exclude frames with high switch rates ($> 0.7$) to ignore abnormal situations such as set-pieces. Consequently, discrete g-segmentation returns an estimated change-point among the sequence of the remaining \emph{valid} adjacency matrices using the pairwise Manhattan distances.

After finding a change-point during the given period, the algorithm decides its significance based on three conditions. To be specific, we recognize the estimated change-point $\tau$ as significant only if (1) the $p$-value of the scan statistic is less than 0.01, (2) both of the segments last for at least 5 minutes, and (3) the mean role-adjacency matrices calculated from the respective segments are far enough (i.e., have large Manhattan distance) from each other. The threshold duration for (2) comes from the fact that abnormal periods such as injury breaks or VAR checks generally last for 2--3 minutes, and therefore 5 minutes are robust against them. Also, the threshold distance for (3) is empirically set to 7.0.

Since there can be more than one formation change-point during a session, we construct a recursive framework to find multiple change-points. First, if there is a significant change-point in the given period, we respectively apply the CPD algorithm to the sequences before and after the change-point again. Each branch-CPD terminates when there is no significant change-point in the segment of interest. In turn, the given session $T$ is partitioned into several formation periods $T_1 < \cdots < T_m$.

\subsection{Formation Clustering Based on Mean Role-Adjacency Matrices} \label{formcpd_cluster}
We now represent the formation in a formation period $T_i$ as a weighted graph $F(T_i) = (V(T_i), A(T_i))$ with the \emph{mean role locations}
\begin{equation}
    V(T_i) =
    \frac{1}{|T_i^*|} \sum_{t \in T_i^*} V(t)
\end{equation}
as vertices and the \emph{mean role-adjacency matrix}
\begin{equation}
    A(T_i) =
    \frac{1}{|T_i^*|} \sum_{t \in T_i^*} A(t)
\end{equation}
as edge matrix, where $T_i^*$ denotes the set of valid time stamps in $T_i$ with small switch rates and $V(t) = (v_1(t), \ldots, v_N(t))^T \in \mathbb{R}^{N \times 2}$ is the 2D locations of $N$ roles normalized to have zero mean at each frame $t$ (i.e., $\sum_{k=1}^N v_k(t) = (0,0)$). See Fig.~\ref{fig:mean_mat} as an example.

Unlike Section~\ref{formcpd_gseg}, we first need to align the roles when calculating the distance between a pair of formation graphs since each role label from 1 to $N$ has nothing to do with the same label in another graph (especially that from another match). Thus, for a pair of formation graphs, we find the ``optimal'' permutation of the role labels of one graph relative to the other and calculate the Manhattan distance (Eq.~\ref{eq:manhattan}) between the adjacency matrices.

To be specific, let $F = (V, A)$ and $F' = (V', A')$ be graphs from a pair of formation periods. Then, we rearrange the rows and columns of $A$ based on the Hungarian algorithm~\citep{Kuhn1955} using the pairwise Euclidean distances between $V$ and $V'$. That is, we find the optimal permutation matrix $Q$ minimizing the assignment cost
\begin{equation*}
    c(V, V'; Q) = \sum_{k=1}^N \|(QV)_k - v'_k\|_2
\end{equation*}
and use $QAQ^T$ and $A'$ when calculating the distance between the formations $F$ and $F'$, i.e.,
\begin{equation}
    d(F,F') = d_M(QAQ^T, A')
\end{equation}

\begin{figure}[htb]
    \centering
    \begin{subfigure}[t]{0.23\textwidth}
        \includegraphics[width=\textwidth]{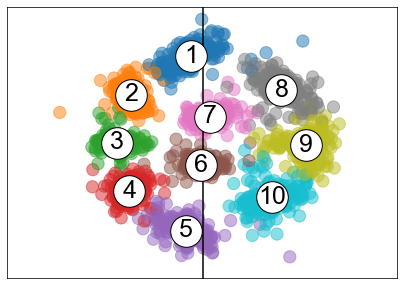}
        \caption{3-4-3 (22.3\%)}
        \label{fig:formation_343}
    \end{subfigure}
    \begin{subfigure}[t]{0.23\textwidth}
        \centering
        \includegraphics[width=\textwidth]{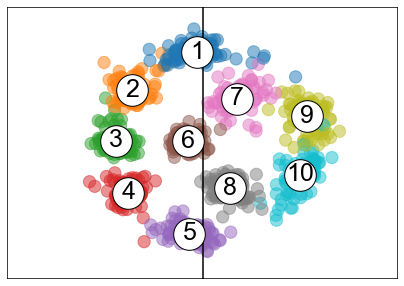}
        \caption{3-5-2 (12.2\%)}
        \label{fig:formation_352}
    \end{subfigure}
    \begin{subfigure}[t]{0.23\textwidth}
        \includegraphics[width=\textwidth]{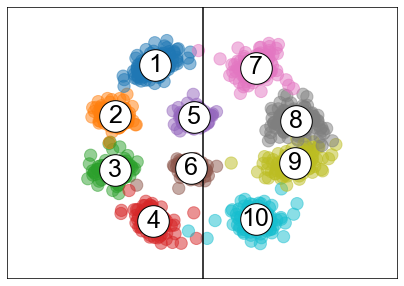}
        \caption{4-4-2 (16.3\%)}
        \label{fig:formation_442}
    \end{subfigure}
    \begin{subfigure}[t]{0.23\textwidth}
        \centering
        \includegraphics[width=\textwidth]{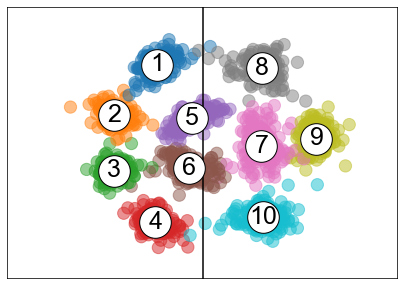}
        \caption{4-2-3-1 (20.9\%)}
        \label{fig:formation_4231}
    \end{subfigure}
    \begin{subfigure}[t]{0.23\textwidth}
        \includegraphics[width=\textwidth]{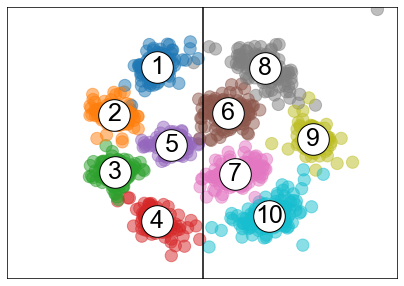}
        \caption{4-3-3 (17.4\%)}
        \label{fig:formation_433}
    \end{subfigure}
    \begin{subfigure}[t]{0.23\textwidth}
        \centering
        \includegraphics[width=\textwidth]{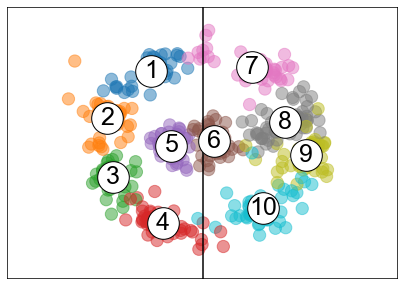}
        \caption{4-1-3-2 (5.8\%)}
        \label{fig:formation_4132}
    \end{subfigure}
    \caption{Mean role locations of each formation group with the proportion (\%) in terms of playing time.}
\label{fig:clusters}
\end{figure}

Along with this distance metric, we cluster the formation obtained from the entire dataset as in Bialkowski et al.~\cite{Bialkowski2014-Large} to decide which formations are the same or different. For this work, we use the GPS data collected from the two seasons (2019 and 2020) of K League 1 and 2, the first two divisions of the South Korean professional soccer league. The data from 809 sessions (i.e., match halves) with at least one moment when ten outfield players were simultaneously measured are split into 864 formation periods and 2,152 role periods. The two-step agglomerative clustering task applied on the pairwise formation distances divides the 864 formations into six main formation groups and ``others'' group consisting of several clusters with less than 15 observations. (Appendix~\ref{formcpd_cluster_details} describes the detailed process.) Lastly, we give each of the six groups a label commonly used by the domain participants, as in Fig.~\ref{fig:clusters}.

The difference of our clustering task from that in Bialkowski et al.~\cite{Bialkowski2014-Large} is the use of the distance between mean role-adjacency matrices instead of the earth mover's distance (EMD) between role locations. While formations in the previous work are from halves of matches with almost the same duration, ours are from formation periods with shorter and varying durations causing distortion of formation graphs. Using mean role-adjacency matrices is more robust to this distortion since they are affected by the adjacency relationships rather than absolute locations.

\section{Role Change-Point Detection} \label{rolecpd}
The process of RoleCPD is analogous to those of FormCPD, except that it leverages role permutations (calculated in Section~\ref{rolecpd_seq}) instead of role-adjacency matrices. Again, discrete g-segmentation recursively finds role change-points in each formation period (Section~\ref{rolecpd_gseg}). Lastly, we give domain-specific position labels to individual players per role period (Section~\ref{role_labeling}).

\begin{table}
  \caption{Assignment rules of position labels to roles.}
  \label{tab:positions}
  \begin{tabular}{ccccccc}
    \toprule
    Role & 3-4-3 & 3-5-2 & 4-4-2 & 4-2-3-1 & 4-3-3 & 4-1-3-2 \\
    \midrule
    1 & LWB & LWB & LB & LB & LB & LB \\
    2 & LCB & LCB & LCB & LCB & LCB & LCB \\
    3 & CB & CB & RCB & RCB & RCB & RCB \\
    4 & RCB & RCB & RB & RB & RB & RB \\
    5 & RWB & RWB & LCM & LDM & CDM & CDM \\
    6 & RCM & CDM & RCM & RDM & LCM & CAM\\
    7 & LCM & LCM & LM & CAM & RCM & LM \\
    8 & LM & RCM & LCF & LM & LM & LCF \\
    9 & CF & LCF & RCF & CF & CF & RCF \\
    10 & RM & RCF & RM & RM & RM & RM \\ 
    \bottomrule
  \end{tabular}
\end{table}

\subsection{Calculating the Sequence of Role Permutations} \label{rolecpd_seq}
During the initialization step of role representation in Section~\ref{formcpd_seq}, each player $p$ is assigned to an \emph{initial role} $X_p$ according to a canonical order such as uniform number. Then, one can express the temporary role assignment to the players $p_1, \ldots, p_N$ at time $t$ as a permutation $\pi_t = (\pi_t(X_{p_1}) \ \cdots \ \pi_t(X_{p_N})) \in \mathrm{S}(\mathcal{X})$  of the initial role assignment $(X_{p_1} \ \cdots \ X_{p_N})$, i.e.,
\begin{equation}
    \beta_t(p) = \pi_t(X_p).
    \label{eq:beta}
\end{equation}

\subsection{Applying Discrete g-Segmentation to the Sequence of Permutations} \label{rolecpd_gseg}
Similar to Section~\ref{formcpd_gseg}, we apply the discrete g-segmentation to the sequence of role permutations $\{ \pi_t \}_{t \in T_i}$ obtained from each formation period $T_i$. Here the algorithm finds an estimated change-point among the sequence of valid permutations (i.e., with switch rates $\le 0.7$) using the Hamming distance
\begin{equation}
    d_H(\pi_t, \pi_{t'})
    = |\{ X : \pi_t(X) \neq \pi_{t'}(X), \ X \in \mathcal{X} \}|
    \label{eq:hamming}
\end{equation}
as the distance measure between role permutations.

Testing the significance of the change-point also emulates that in FormCPD except for the last condition. Since the goal of RoleCPD is to find the point when the dominant role assignment changes, we recognize the change-point $\tau$ as significant only if the most frequent permutations differ between before and after $\tau$.

Finally, the recursive CPD applied to the sequence from each formation period $T_i$ results in a partition $T_{i,1} < \cdots < T_{i,n_i}$ of $T_i$. We set the instructed role per player in each role period $T_{i,j}$ be the most frequent permutation, and express every temporary roles resulting from the role representation as a permutation of the instructed roles. Formally speaking, we define the P-IR maps $\{ \alpha_t: P \rightarrow \mathcal{X} \}_{t \in T_{i,j}}$ as
\begin{equation}
    \alpha_t(p) = \pi_{(i,j)}(X_p) \quad
    \text{(constant along $t \in T_{i,j}$)}
    \label{eq:alpha}
\end{equation}
where $\pi_{(i,j)} \in S(\mathcal{X})$ is the most frequent permutation among $\{ \pi_t \}_{t \in T_{i,j}}$. The RolePerm $\sigma_t(X_p) = \beta_t \circ \alpha_t^{-1}$ at $t \in T_{i,j}$ can then be obtained as
\begin{equation}
    \sigma_t(X_p) = \beta_t(\alpha_t^{-1}(X_p)) = \pi_t(\pi_{(i,j)}^{-1}(X_p))
    \label{eq:sigma}
\end{equation}
from Eq.~\ref{eq:beta} and Eq.~\ref{eq:alpha}.

The resulting P-IR maps satisfy the three conditions presented in Section~\ref{rolecpd_def}. They satisfy the period-wise consistency and uniqueness by Eq.~\ref{eq:alpha} (i.e., $\alpha_t(p)$ is constant along $t \in T_{i,j}$ and distinct across $p \in P$ since it is a permutation of distinct initial roles). Also, the aforementioned significance test guarantees the existence of a role change between adjacent role periods.

\subsection{Assigning Domain-Specific Position Labels to Instructed Roles of Players} \label{role_labeling}
Since roles are independently initialized based on the players' uniform numbers per session in Section~\ref{rolecpd_seq}, each role label from 1 to 10 in a formation graph has nothing to do with the same label in another graph. Thus, we first align role labels between formation graphs so that the locations of the same label from different graphs are close to one another. Then, we give one of the 18 position labels such as ``left wing-back (LWB)'' or ``central defensive midfielder (CDM)'' to each of the aligned roles 1--10 per formation group based on domain knowledge.

The role alignment is achieved by clustering the roles in each formation group based on the EM algorithm similar to the role representation~\cite{Bialkowski2014-Large} as follows:
\begin{itemize}
    \item Initialization: Partition the role locations in each formation group into ten clusters by agglomerative clustering.
    \item E-step: Apply the Hungarian algorithm~\citep{Kuhn1955} between the ten role locations and the cluster centroids (mean locations) per formation graph to reassign cluster labels to roles.
    \item M-step: Update the centroid per cluster.
\end{itemize}
As a result, roles are relabeled as the annotations in Fig.~\ref{fig:clusters} so that the same label in the same formation graph corresponds to a certain ``position'' in the pitch.

\begin{figure}
    \centering
    \includegraphics[width=0.47\textwidth]{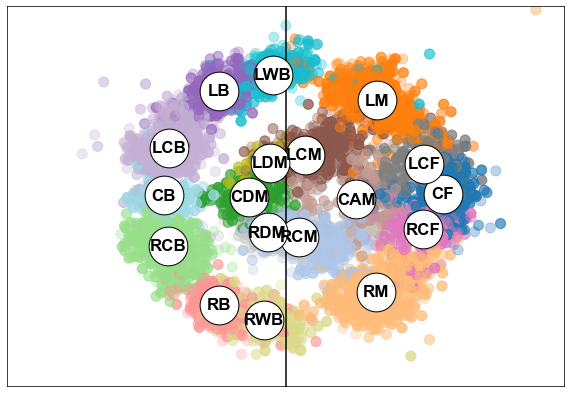}
    \caption{Role locations in the ordinary formation groups colored by position label.}
    \label{fig:pos_scatter}
\end{figure}

\begin{figure}[htb]
    \centering
    \begin{subfigure}[t]{0.23\textwidth}
        \includegraphics[width=\textwidth]{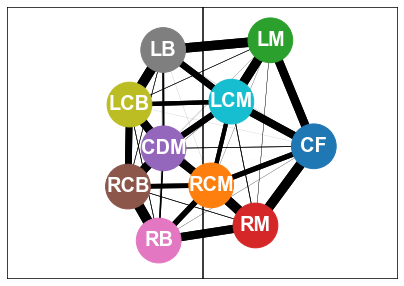}
        \caption{Formation ``4-3-3''}
        \label{fig:pos_annot_normal}
    \end{subfigure}
    \begin{subfigure}[t]{0.23\textwidth}
        \centering
        \includegraphics[width=\textwidth]{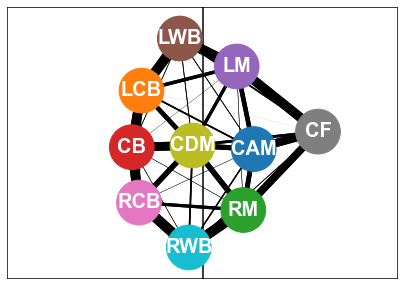}
        \caption{Formation ``others''}
        \label{fig:pos_annot_others}
    \end{subfigure}
    \caption{Examples of position-labeled formation graph.}
    \label{fig:pos_annot}
\end{figure}

As the last step, we give ten position labels to the instructed roles 1--10 per formation group by the rules in Table~\ref{tab:positions}. This results in each player having one position label among the total of 18 in Fig.~\ref{fig:pos_scatter} per role period. For the ``others'' formation group, we assign position labels separately for each formation graph based on the Hungarian algorithm~\citep{Kuhn1955} between the ten role locations and the mean locations of the 18 position labels obtained from the above. Fig.~\ref{fig:pos_annot} shows the resultant position-labeled graphs as examples.

\begin{figure*}[t]
    \centering
    \begin{subfigure}[t]{0.36\textwidth}
        \includegraphics[width=\textwidth]{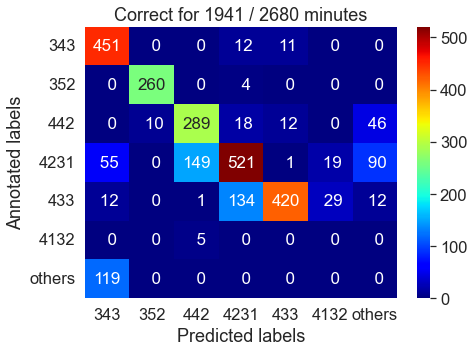}
        \caption{Formation accuracy in minutes}
        \label{fig:form_conf_mat}
    \end{subfigure}
    \begin{subfigure}[t]{0.6\textwidth}
        \centering
        \includegraphics[width=\textwidth]{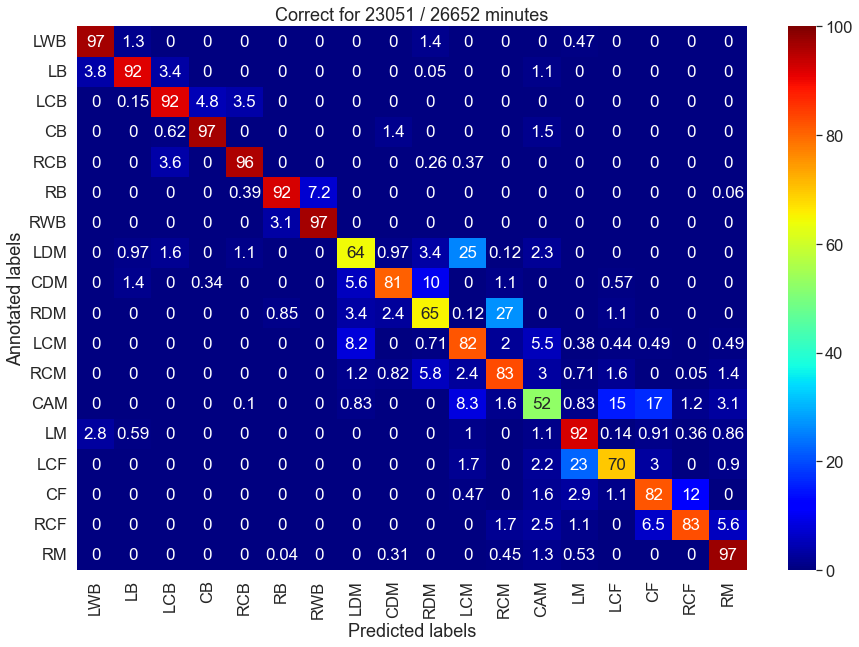}
        \caption{Position accuracy in percentage}
        \label{fig:role_conf_mat}
    \end{subfigure}
    \caption{Confusion matrices for formation and position predictions. Note that we calculate the matrix in minutes and in percentage for Fig~\ref{fig:form_conf_mat} and Fig~\ref{fig:role_conf_mat}, respectively, since the former has high class imbalance while the latter does not.}
    \label{fig:conf_mat}
\end{figure*}

\section{Experiments}
In this section, we first describe the evaluation process of SoccerCPD by comparing our results with annotations by domain experts (Section \ref{model_eval}). Next, we introduce switching pattern discovery (Section \ref{switch_discovery}) and set-piece detection (Section \ref{setpiece}), which are useful and interpretable applications of our model.

\subsection{Model Evaluation} \label{model_eval}
We evaluated the performance of SoccerCPD by measuring the prediction accuracies of (1) team formation and (2) player position. Namely, we calculated the ratios of correctly detected one-minute segments to the total number of segments (total minutes played).

More specifically, we used the ground truth labels annotated by domain experts who worked as coaches or video analysts in elite soccer teams. We divided the whole playing time into one-minute segments and compared the predicted and annotated formation/position labels per segment. (For instance, there are 90 formation annotations and 900 position annotations for a single team's 90-minute match without a red card or missing measurement.) As the minute-wise annotation of formation and role is very costly, we sampled 28 matches, one for each of 28 season-team pairs (i.e., 15 teams in season 2019 and 13 in 2020), and let the experts perform the annotation only for those matches.

Fig.~\ref{fig:conf_mat} demonstrates the detailed results in confusion matrices. For the formation prediction, our result accords with the annotation for 1,941 minutes (72.4\%) among 2,680 minutes of the total playing time. By analyzing the failure cases, we have found that our method can make a misprediction in the following patterns:
\begin{itemize}
    \item It sometimes confuses the formations 4-4-2 and 4-2-3-1 because of their similarity. (See the (4,3)-element of the confusion matrix in Fig.~\ref{fig:form_conf_mat}.) Particularly, CAM and CF sometimes form a horizontal line when the team is defending, making the prediction more difficult.
    \item It cannot distinguish 3-4-1-2 (included in the formation ``others'' in our experiment) from 3-4-3, which leads to the (6,1)-element in Fig.~\ref{fig:form_conf_mat}. 3-4-1-2 deploys LCF-CAM-RCF in their forward line, but their arrangement is too similar to the LM-CF-RM line to be separated from 3-4-3.
\end{itemize}

For the position prediction, since labels are compared player-by-player in each segment, the accuracy is calculated for the 26,652 minutes of player-segment pairs in total (i.e., about ten times the total playing time). Overall, the predicted labels accord with the annotated ones for 23,051 minutes (86.5\%). Most positions achieve recalls higher than 80\%, except for LDM, RDM, CAM, and LCF whose low recalls are caused by the aforementioned mispredictions of formations. This result is far greater than the formation prediction, implying that our method finds most of the positions correctly even when it fails to predict a formation.



\subsection{Switching Pattern Discovery} \label{switch_discovery}
The RolePerms obtained from Eq.~\ref{eq:sigma} indicate the temporary role switches between players. If every player maintains the originally instructed role at time $t$, then the RolePerm $\sigma_t$ becomes the identity permutation. In other words, a non-identity RolePerm implies that there is a switching play at that time. Thus, we can identify teams' playing patterns by analyzing the non-identity RolePerms.

\begin{figure*}[!htb]
    \centering
    \begin{tabular}{cc}
        \begin{subfigure}[b]{0.46\textwidth}
            \includegraphics[width=\textwidth]{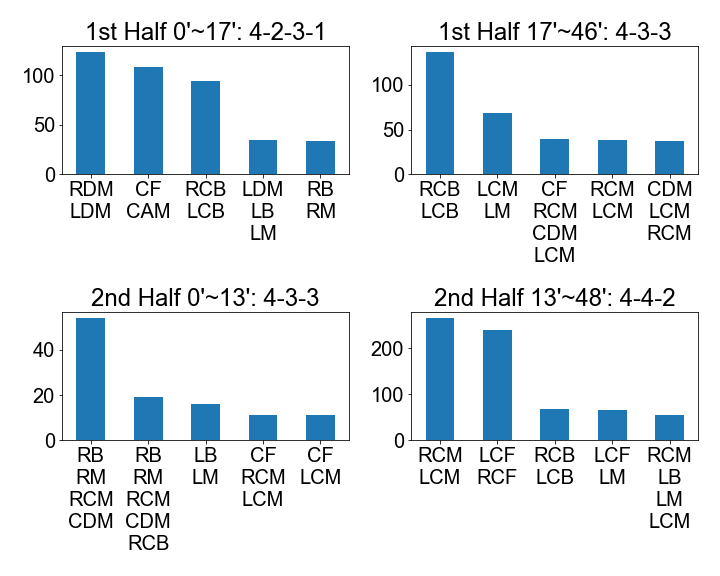}
            \caption{Histograms of top-5 frequent cycles in second.}
            \label{fig:switch_hist}
            \smallskip
        \end{subfigure}
        &
        \begin{tabular}[b]{cc}
        \smallskip
            \begin{subfigure}[t]{0.23\textwidth}
                \centering
                \includegraphics[width=\textwidth]{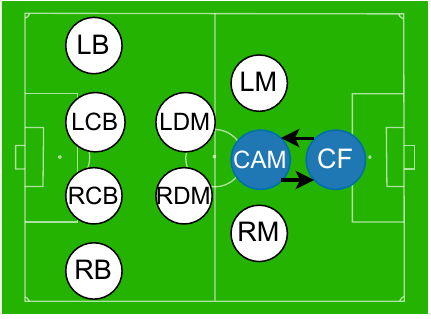}
                \caption{False-nine play in FP-1}
                \label{fig:switch_false9_4231}
            \end{subfigure} &
            \begin{subfigure}[t]{0.23\textwidth}
                \centering
                \includegraphics[width=\textwidth]{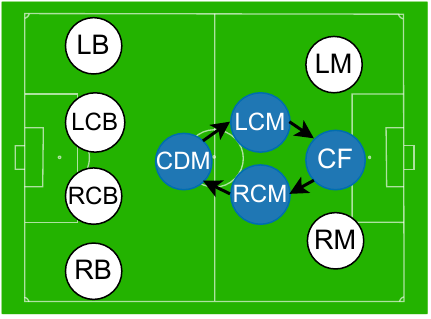}
                \caption{False-nine play in FP-2}
                \label{fig:switch_false9_433}
            \end{subfigure}
        \\ \smallskip
            \begin{subfigure}[t]{0.23\textwidth}
                \centering
                \includegraphics[width=\textwidth]{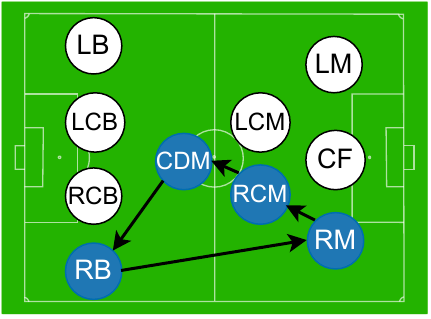}
                \caption{Fullback overlap in FP-3}
                \label{fig:switch_overlap}
            \end{subfigure} &
            \begin{subfigure}[t]{0.23\textwidth}
                \centering
                \includegraphics[width=\textwidth]{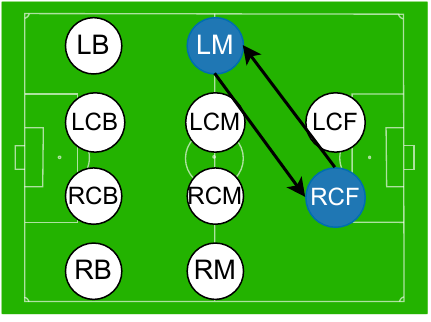}
                \caption{Cutting inside in FP-4}
                \label{fig:switch_cutin}
            \end{subfigure}
        \end{tabular}
    \end{tabular}
    \caption{(a) Frequent cycles per formation period, and (b)--(e) some tactically notable cycles from the match in Fig.~\ref{fig:timeline}.}
\end{figure*}

\begin{figure*}
    \centering
    \includegraphics[width=0.9\textwidth]{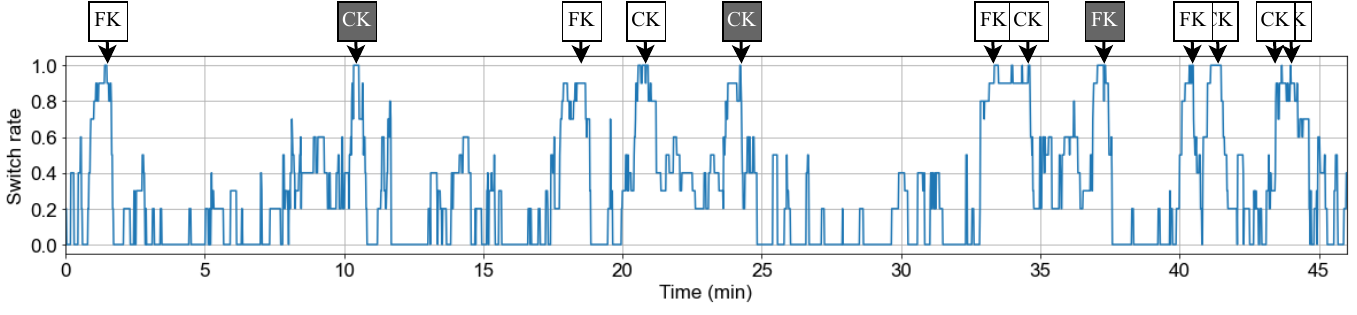}
    \caption{A time-series of switch rates and set-piece occurrence in the first half of the match in Fig.~\ref{fig:timeline}. Squares with the labels ``CK'' and ``FK'' mean a corner kick and a close free kick (that the ball was kicked into the box), respectively. The color of a square indicates the attacking team during the situation, where the white corresponds to the measured team of interest.}
\label{fig:setpieces}
\end{figure*}

A point to note is that independent switches can occur in the different parts of the pitch at the same time. Thus, we decompose each RolePerm into one or multiple cyclic permutations (which we call \emph{cyclic RolePerms} or simply \emph{cycles}). For example, the RolePerm
\begin{equation*}
    \sigma =
    \left(
        \begin{array}{ccccccc}
        \text{LB} & \cdots & \text{LM} & \cdots & \text{RM} & \cdots & \text{RCF} \\
        \text{LM} & \cdots & \text{LB} & \cdots & \text{RCF} & \cdots & \text{RM}
        \end{array}
    \right)
\end{equation*}
with switch rate 0.4 is the concurrence of the two 2-cycles (LB LM) and (RM RCF). Here, LB and RCF raise a non-identity RolePerm $\sigma$ together, but they did not switch roles with each other. Considering this kind of cases, we regard concurrent cycles as distinct role switches to solely focus on the ``interchanging'' patterns of roles.

As a concrete case study, we look into the cycles during the match introduced in Fig.~\ref{fig:timeline} and derive some domain insights. Fig.~\ref{fig:switch_hist} shows the top-5 frequent cycles with position labels per formation period. For example, the leftmost bar in the first histogram means that RDM and LDM switched with each other for 120 seconds in formation period 1. Notable observations are as follows, where we abbreviate formation/role periods as FPs/RPs.
\begin{itemize}
    \item The players dynamically switched their roles in FP-1 and FP-4 (top-5 cycles lasted for about 400 and 750 seconds, respectively), while there were few interchanges in FP-2 and FP-3 (only with 280 and 100 seconds, respectively).
    \item The center forward performed the ``false-nine play'' by dropping deep into the midfield positions, generating (CF CAM) in FP-1, (CF RCM CDM LCM) in FP-2, (CF RCM LCM) and (CF LCM) in FP-3.
    \item There were fewer switches in FP-3 by and large, but the overlapping of the right back stood out with the cycles (RB RM RCM CDM) and (RB RM RCM CDM RCB).
    \item In FP-4, the two forwards in 4-4-2 frequently switched with each other. In particular, the cycle (RCF LM) indicates that the left midfielder kept ``cutting inside'' the box.
\end{itemize}

\subsection{Set-Piece Detection} \label{setpiece}
The ``set-piece'' in soccer refers to situations such as corner kicks and free kicks that resume the match from a stoppage by kicking the ball into the scoring zone. Since set-pieces are a great opportunity to score a goal \citep{Power2018}, teams carefully review set-piece situations from previous matches and set aside special tactics for set plays.

As a byproduct of SoccerCPD, we can also help soccer teams easily extract and manage the set play data by automatically detecting set-pieces based on the statistic \emph{switch rate}. The term switch rate has been defined in Section~\ref{formcpd_gseg}, but we slightly alter its definition to the Hamming distance (Eq.~\ref{eq:hamming}) between the RolePerm and the identity permutation divided by the number of roles. (Note that we cannot use this definition in Section~\ref{formcpd_gseg}, as it is before obtaining the RolePerms.) Since the players are totally mixed up during set-pieces, switch rates in this situation soar close to 1.0. Thus, we can construct a simple, totally unsupervised, but fairly accurate set-piece detection model that picks situations with high switch rates (such as $\ge 0.9$). Figure~\ref{fig:setpieces} shows the strong co-occurrence pattern of set-pieces and high switch rates.

\subsection{Discussion on Limitations}
While SoccerCPD offers benefits such as scalability and usability by only relying on player locations, it leaves some limitations as below. For instance, a team can take separate formations for attacking and defending situations, but our framework does not take these contexts into account. Also, ours tends to confuse the formations with similar spatial configurations, since it does not consider on-the-ball actions. The mispredictions of 3-4-1-2 to 3-4-3 and 4-2-3-1 to 4-4-2 in Section~\ref{model_eval} are cases in point.

Another limitation is that our framework only predicts popular formations well. In other words, since our framework gives formation labels only for the clusters with enough sizes, it just classifies irregular formations into ``others'' group. This brings about time delay to reflect the tactics of some adventurous managers, such as asymmetrical formations by Pep Guardiola or Jos\'e Mourinho.

\section{Conclusion}
This study proposes a change-point detection (CPD) framework named SoccerCPD that distinguishes tactically intended formation and role changes from temporary changes in soccer matches. First, we express the temporary role topology and transposition as sequences of binary matrices and permutations, respectively. Considering that these representations are high-dimensional or non-Euclidean data with frequent repetitions, we then apply a graph-based CPD algorithm named discrete g-segmentation to find formation and role assignment change-points.

Since the concept of formation and role is the most fundamental and intuitive way to express soccer tactics, tracking and summarizing the team's formation/role changes has its own worth for domain participants. 
In addition, one can also use additional information such as temporary permutations to discover the switching patterns or to detect set-pieces as introduced in our study. Hence, we expect our method to be widespread in the sports industry.

\begin{acks}
This work was supported by the National Research Foundation of Korea (NRF) grant funded by the Korean government (MSIT) (No. 2020R1F1A1073451 and No. 2020R1A4A3079947). Also, the authors thank Taein Lee, Minwoo Seo, and Donghwa Shin at Fitogether Inc. for supporting the model evaluation task by annotating formation and position labels as domain experts.
\end{acks}

\bibliographystyle{ACM-Reference-Format}
\bibliography{soccercpd}


\begin{thebibliography}{33}


\ifx \showCODEN    \undefined \def \showCODEN     #1{\unskip}     \fi
\ifx \showDOI      \undefined \def \showDOI       #1{#1}\fi
\ifx \showISBNx    \undefined \def \showISBNx     #1{\unskip}     \fi
\ifx \showISBNxiii \undefined \def \showISBNxiii  #1{\unskip}     \fi
\ifx \showISSN     \undefined \def \showISSN      #1{\unskip}     \fi
\ifx \showLCCN     \undefined \def \showLCCN      #1{\unskip}     \fi
\ifx \shownote     \undefined \def \shownote      #1{#1}          \fi
\ifx \showarticletitle \undefined \def \showarticletitle #1{#1}   \fi
\ifx \showURL      \undefined \def \showURL       {\relax}        \fi
\providecommand\bibfield[2]{#2}
\providecommand\bibinfo[2]{#2}
\providecommand\natexlab[1]{#1}
\providecommand\showeprint[2][]{arXiv:#2}

\bibitem[Aminikhanghahi and Cook(2017)]%
        {Aminikhanghahi2017}
\bibfield{author}{\bibinfo{person}{Samaneh Aminikhanghahi} {and}
  \bibinfo{person}{Diane~J. Cook}.} \bibinfo{year}{2017}\natexlab{}.
\newblock \showarticletitle{{A survey of methods for time series change point
  detection}}.
\newblock \bibinfo{journal}{\emph{Knowledge and Information Systems}}
  \bibinfo{volume}{51} (\bibinfo{year}{2017}), \bibinfo{pages}{339--367}.
\newblock
\showISBNx{2163684814}


\bibitem[Bai(1994)]%
        {Bai1994}
\bibfield{author}{\bibinfo{person}{Jushan Bai}.}
  \bibinfo{year}{1994}\natexlab{}.
\newblock \showarticletitle{{Least squares estimation of a shift in linear
  processes}}.
\newblock \bibinfo{journal}{\emph{Journal of Time Series Analysis}}
  \bibinfo{volume}{15}, \bibinfo{number}{5} (\bibinfo{year}{1994}),
  \bibinfo{pages}{453--472}.
\newblock
\showISSN{0143-9782}


\bibitem[Bai(1997)]%
        {Bai1997}
\bibfield{author}{\bibinfo{person}{Jushan Bai}.}
  \bibinfo{year}{1997}\natexlab{}.
\newblock \showarticletitle{{Estimation of a change point in multiple
  regression models}}.
\newblock \bibinfo{journal}{\emph{Review of Economics and Statistics}}
  \bibinfo{volume}{79}, \bibinfo{number}{4} (\bibinfo{year}{1997}),
  \bibinfo{pages}{551--563}.
\newblock
\showISSN{0034-6535}


\bibitem[Bialkowski et~al\mbox{.}(2014a)]%
        {Bialkowski2014-Win}
\bibfield{author}{\bibinfo{person}{Alina Bialkowski}, \bibinfo{person}{Patrick
  Lucey}, \bibinfo{person}{Peter Carr}, \bibinfo{person}{Yisong Yue}, {and}
  \bibinfo{person}{Iain Matthews}.} \bibinfo{year}{2014}\natexlab{a}.
\newblock \showarticletitle{{“Win at home and draw away”: Automatic
  formation analysis highlighting the differences in home and away team
  behaviors}}. In \bibinfo{booktitle}{\emph{MIT Sloan Sports Analytics
  Conference}}.
\newblock


\bibitem[Bialkowski et~al\mbox{.}(2014b)]%
        {Bialkowski2014-Identifying}
\bibfield{author}{\bibinfo{person}{Alina Bialkowski}, \bibinfo{person}{Patrick
  Lucey}, \bibinfo{person}{Peter Carr}, \bibinfo{person}{Yisong Yue},
  \bibinfo{person}{Sridha Sridharan}, {and} \bibinfo{person}{Iain Matthews}.}
  \bibinfo{year}{2014}\natexlab{b}.
\newblock \showarticletitle{{Identifying team style in soccer using formations
  learned from spatiotemporal tracking data}}. In
  \bibinfo{booktitle}{\emph{IEEE International Conference on Data Mining
  Workshops}}.
\newblock


\bibitem[Bialkowski et~al\mbox{.}(2014c)]%
        {Bialkowski2014-Large}
\bibfield{author}{\bibinfo{person}{Alina Bialkowski}, \bibinfo{person}{Patrick
  Lucey}, \bibinfo{person}{Peter Carr}, \bibinfo{person}{Yisong Yue},
  \bibinfo{person}{Sridha Sridharan}, {and} \bibinfo{person}{Iain Matthews}.}
  \bibinfo{year}{2014}\natexlab{c}.
\newblock \showarticletitle{{Large-scale analysis of soccer matches using
  spatiotemporal tracking data}}. In \bibinfo{booktitle}{\emph{IEEE
  International Conference on Data Mining}}.
\newblock
\showISBNx{978-1-4799-4302-9}


\bibitem[Chen and Zhang(2015)]%
        {Chen2015}
\bibfield{author}{\bibinfo{person}{Hao Chen} {and} \bibinfo{person}{Nancy
  Zhang}.} \bibinfo{year}{2015}\natexlab{}.
\newblock \showarticletitle{{Graph-based change-point detection}}.
\newblock \bibinfo{journal}{\emph{Annals of Statistics}} \bibinfo{volume}{43},
  \bibinfo{number}{1} (\bibinfo{year}{2015}), \bibinfo{pages}{139--176}.
\newblock
\showISSN{00905364}


\bibitem[Chu and Chen(2019)]%
        {Chu2019}
\bibfield{author}{\bibinfo{person}{Lynna Chu} {and} \bibinfo{person}{Hao
  Chen}.} \bibinfo{year}{2019}\natexlab{}.
\newblock \showarticletitle{{Asymptotic distribution-free change-point
  detection for multivariate and non-euclidean data}}.
\newblock \bibinfo{journal}{\emph{Annals of Statistics}} \bibinfo{volume}{47},
  \bibinfo{number}{1} (\bibinfo{year}{2019}), \bibinfo{pages}{382--414}.
\newblock
\showISSN{23318422}


\bibitem[Delaunay(1934)]%
        {Delaunay1934}
\bibfield{author}{\bibinfo{person}{Boris Delaunay}.}
  \bibinfo{year}{1934}\natexlab{}.
\newblock \showarticletitle{{Sur la sph{\`{e}}re vide}}.
\newblock \bibinfo{journal}{\emph{Bulletin de l'Acad{\'{e}}mie des Sciences de
  l'URSS, Classe des Sciences Math{\'{e}}matiques et Naturelles}}
  \bibinfo{volume}{6} (\bibinfo{year}{1934}), \bibinfo{pages}{793--800}.
\newblock


\bibitem[Gretton et~al\mbox{.}(2012)]%
        {Gretton2012}
\bibfield{author}{\bibinfo{person}{Arthur Gretton}, \bibinfo{person}{Karsten~M.
  Borgwardt}, \bibinfo{person}{Malte~J. Rasch}, \bibinfo{person}{Bernhard
  Sch{\"{o}}lkopf}, {and} \bibinfo{person}{Alexander Smola}.}
  \bibinfo{year}{2012}\natexlab{}.
\newblock \showarticletitle{{A kernel two-sample test}}.
\newblock \bibinfo{journal}{\emph{Journal of Machine Learning Research}}
  \bibinfo{volume}{13} (\bibinfo{year}{2012}), \bibinfo{pages}{723--773}.
\newblock
\showISSN{15324435}


\bibitem[Harchaoui et~al\mbox{.}(2009)]%
        {Harchaoui2009}
\bibfield{author}{\bibinfo{person}{Za{\"{i}}d Harchaoui},
  \bibinfo{person}{Francis Bach}, {and} \bibinfo{person}{{\'{E}}ric Moulines}.}
  \bibinfo{year}{2009}\natexlab{}.
\newblock \showarticletitle{{Kernel change-point analysis}}.
\newblock \bibinfo{journal}{\emph{Advances in Neural Information Processing
  Systems}} (\bibinfo{year}{2009}), \bibinfo{pages}{609--616}.
\newblock
\showISBNx{9781605609492}


\bibitem[Kim et~al\mbox{.}(2021)]%
        {Kim2021}
\bibfield{author}{\bibinfo{person}{Hyunsung Kim}, \bibinfo{person}{Jihun Kim},
  \bibinfo{person}{Dongwook Chung}, \bibinfo{person}{Jonghyun Lee},
  \bibinfo{person}{Jinsung Yoon}, {and} \bibinfo{person}{Sang-Ki Ko}.}
  \bibinfo{year}{2021}\natexlab{}.
\newblock \showarticletitle{{6MapNet : Representing soccer players from
  tracking data by a triplet network}}. In \bibinfo{booktitle}{\emph{ECML-PKDD
  Workshop on Machine Learning and Data Mining for Sports Analytics}}.
\newblock


\bibitem[Ko et~al\mbox{.}(2015)]%
        {Ko2015}
\bibfield{author}{\bibinfo{person}{Stanley~I.M. Ko},
  \bibinfo{person}{Terence~T.L. Chong}, {and} \bibinfo{person}{Pulak Ghosh}.}
  \bibinfo{year}{2015}\natexlab{}.
\newblock \showarticletitle{{Dirichlet process hidden Markov multiple
  change-point model}}.
\newblock \bibinfo{journal}{\emph{Bayesian Analysis}} \bibinfo{volume}{10},
  \bibinfo{number}{2} (\bibinfo{year}{2015}), \bibinfo{pages}{275--296}.
\newblock
\showISSN{19316690}


\bibitem[Kuhn(1955)]%
        {Kuhn1955}
\bibfield{author}{\bibinfo{person}{H.~W. Kuhn}.}
  \bibinfo{year}{1955}\natexlab{}.
\newblock \showarticletitle{{The Hungarian method for the assignment problem}}.
\newblock \bibinfo{journal}{\emph{Naval Research Logistics Quarterly}}
  \bibinfo{volume}{2}, \bibinfo{number}{1-2} (\bibinfo{year}{1955}),
  \bibinfo{pages}{83--97}.
\newblock
\showISSN{00281441}


\bibitem[Lajugie et~al\mbox{.}(2014)]%
        {Lajugie2014}
\bibfield{author}{\bibinfo{person}{R{\'{e}}mi Lajugie},
  \bibinfo{person}{Sylvain Arlot}, {and} \bibinfo{person}{Francis Bach}.}
  \bibinfo{year}{2014}\natexlab{}.
\newblock \showarticletitle{{Large-margin metric learning for constrained
  partitioning problems}}. In \bibinfo{booktitle}{\emph{International
  Conference on Machine Learning}}. \bibinfo{pages}{502--510}.
\newblock
\showISBNx{9781634393973}


\bibitem[Lavielle and Moulines(2000)]%
        {Lavielle2000}
\bibfield{author}{\bibinfo{person}{Marc Lavielle} {and} \bibinfo{person}{Eric
  Moulines}.} \bibinfo{year}{2000}\natexlab{}.
\newblock \showarticletitle{{Least‐squares estimation of an unknown number of
  shifts in a time series}}.
\newblock \bibinfo{journal}{\emph{Journal of Time Series Analysis}}
  \bibinfo{volume}{21}, \bibinfo{number}{1} (\bibinfo{year}{2000}),
  \bibinfo{pages}{33--59}.
\newblock
\showISBNx{9781626239777}
\showISSN{0143-9782}


\bibitem[Lung-Yut-Fong et~al\mbox{.}(2015)]%
        {Lung-Yut-Fong2015}
\bibfield{author}{\bibinfo{person}{Alexandre Lung-Yut-Fong},
  \bibinfo{person}{C{\'{e}}line L{\'{e}}vy-Leduc}, {and}
  \bibinfo{person}{Olivier Capp{\'{e}}}.} \bibinfo{year}{2015}\natexlab{}.
\newblock \showarticletitle{{Homogeneity and change-point detection tests for
  multivariate data using rank statistics}}.
\newblock \bibinfo{journal}{\emph{Journal de la Soci{\'{e}}t{\'{e}}
  Fran{\c{c}}aise de Statistique}} \bibinfo{volume}{156}, \bibinfo{number}{4}
  (\bibinfo{year}{2015}), \bibinfo{pages}{133--162}.
\newblock


\bibitem[Machado et~al\mbox{.}(2017)]%
        {Machado2017}
\bibfield{author}{\bibinfo{person}{Vinicius Machado}, \bibinfo{person}{Roger
  Leite}, \bibinfo{person}{Felipe Moura}, \bibinfo{person}{Sergio Cunha},
  \bibinfo{person}{Filip Sadlo}, {and} \bibinfo{person}{Jo{\~{a}}o~L.D.
  Comba}.} \bibinfo{year}{2017}\natexlab{}.
\newblock \showarticletitle{{Visual soccer match analysis using spatiotemporal
  positions of players}}.
\newblock \bibinfo{journal}{\emph{Computers and Graphics}}
  \bibinfo{volume}{68} (\bibinfo{year}{2017}), \bibinfo{pages}{84--95}.
\newblock
\showISSN{00978493}


\bibitem[Narizuka and Yamazaki(2017)]%
        {Narizuka2017}
\bibfield{author}{\bibinfo{person}{Takuma Narizuka} {and}
  \bibinfo{person}{Yoshihiro Yamazaki}.} \bibinfo{year}{2017}\natexlab{}.
\newblock \showarticletitle{{Characterization of the formation structure in
  team sports}}. In \bibinfo{booktitle}{\emph{Proceedings of the Institute of
  Statistical Mathematics}}, Vol.~\bibinfo{volume}{65}.
  \bibinfo{pages}{299--307}.
\newblock


\bibitem[Narizuka and Yamazaki(2019)]%
        {Narizuka2019}
\bibfield{author}{\bibinfo{person}{Takuma Narizuka} {and}
  \bibinfo{person}{Yoshihiro Yamazaki}.} \bibinfo{year}{2019}\natexlab{}.
\newblock \showarticletitle{{Clustering algorithm for formations in football
  games}}.
\newblock \bibinfo{journal}{\emph{Scientific Reports}}  \bibinfo{volume}{9}
  (\bibinfo{year}{2019}).
\newblock
\showISSN{2045-2322}


\bibitem[Pein et~al\mbox{.}(2017)]%
        {Pein2017}
\bibfield{author}{\bibinfo{person}{Florian Pein}, \bibinfo{person}{Hannes
  Sieling}, {and} \bibinfo{person}{Axel Munk}.}
  \bibinfo{year}{2017}\natexlab{}.
\newblock \showarticletitle{{Heterogeneous change point inference}}.
\newblock \bibinfo{journal}{\emph{Journal of the Royal Statistical Society.
  Series B: Statistical Methodology}} \bibinfo{volume}{79}, \bibinfo{number}{4}
  (\bibinfo{year}{2017}), \bibinfo{pages}{1207--1227}.
\newblock
\showISSN{14679868}


\bibitem[Power et~al\mbox{.}(2018)]%
        {Power2018}
\bibfield{author}{\bibinfo{person}{Paul Power}, \bibinfo{person}{Jennifer
  Hobbs}, \bibinfo{person}{Hector Ruiz}, \bibinfo{person}{Xinyu Wei}, {and}
  \bibinfo{person}{Patrick Lucey}.} \bibinfo{year}{2018}\natexlab{}.
\newblock \showarticletitle{{Mythbusting set-pieces in soccer}}. In
  \bibinfo{booktitle}{\emph{MIT Sloan Sports Analytics Conference}}.
\newblock


\bibitem[Qu and Perron(2007)]%
        {Qu2007}
\bibfield{author}{\bibinfo{person}{Zhongjun Qu} {and} \bibinfo{person}{Pierre
  Perron}.} \bibinfo{year}{2007}\natexlab{}.
\newblock \showarticletitle{{Estimating and testing structural changes in
  multivariate regressions}}.
\newblock \bibinfo{journal}{\emph{Econometrica}} \bibinfo{volume}{75},
  \bibinfo{number}{2} (\bibinfo{year}{2007}), \bibinfo{pages}{459--502}.
\newblock
\showISSN{00129682}


\bibitem[Rein and Memmert(2016)]%
        {Rein2016}
\bibfield{author}{\bibinfo{person}{Robert Rein} {and} \bibinfo{person}{Daniel
  Memmert}.} \bibinfo{year}{2016}\natexlab{}.
\newblock \showarticletitle{{Big data and tactical analysis in elite soccer:
  future challenges and opportunities for sports science}}.
\newblock \bibinfo{journal}{\emph{SpringerPlus}} (\bibinfo{year}{2016}).
\newblock
\showISBNx{1303-2968}
\showISSN{21931801}


\bibitem[Sha et~al\mbox{.}(2016)]%
        {Sha2016}
\bibfield{author}{\bibinfo{person}{Long Sha}, \bibinfo{person}{Patrick Lucey},
  \bibinfo{person}{Yisong Yue}, \bibinfo{person}{Peter Carr},
  \bibinfo{person}{Charlie Rohlf}, {and} \bibinfo{person}{Iain Matthews}.}
  \bibinfo{year}{2016}\natexlab{}.
\newblock \showarticletitle{{Chalkboarding: A new spatiotemporal query paradigm
  for sports play retrieval}}. In \bibinfo{booktitle}{\emph{International
  Conference on Intelligent User Interfaces}}. \bibinfo{pages}{336--347}.
\newblock
\showISBNx{9781450341370}


\bibitem[Shaw and Glickman(2019)]%
        {Shaw2019}
\bibfield{author}{\bibinfo{person}{Laurie Shaw} {and} \bibinfo{person}{Mark
  Glickman}.} \bibinfo{year}{2019}\natexlab{}.
\newblock \showarticletitle{{Dynamic analysis of team strategy in professional
  football}}. In \bibinfo{booktitle}{\emph{Bar{\c{c}}a Sport Analytics
  Summit}}. \bibinfo{pages}{1--13}.
\newblock


\bibitem[Shawe-Taylor and Cristianini(2004)]%
        {Shawe-Taylor2004}
\bibfield{author}{\bibinfo{person}{John Shawe-Taylor} {and}
  \bibinfo{person}{Nello Cristianini}.} \bibinfo{year}{2004}\natexlab{}.
\newblock \bibinfo{booktitle}{\emph{{Kernel Methods for Pattern Analysis}}}.
\newblock \bibinfo{publisher}{Cambridge University Press}.
\newblock


\bibitem[Song and Chen(2020)]%
        {Song2020}
\bibfield{author}{\bibinfo{person}{Hoseung Song} {and} \bibinfo{person}{Hao
  Chen}.} \bibinfo{year}{2020}\natexlab{}.
\newblock \showarticletitle{{Asymptotic distribution-free change-point
  detection for data with repeated observations}}.
\newblock  (\bibinfo{year}{2020}).
\newblock
\showISSN{23318422}
\showeprint[arxiv]{2006.10305}


\bibitem[Tamura and Masuda(2015)]%
        {Tamura2015}
\bibfield{author}{\bibinfo{person}{Kohei Tamura} {and} \bibinfo{person}{Naoki
  Masuda}.} \bibinfo{year}{2015}\natexlab{}.
\newblock \showarticletitle{{Win-stay lose-shift strategy in formation changes
  in football}}.
\newblock \bibinfo{journal}{\emph{EPJ Data Science}} \bibinfo{volume}{4},
  \bibinfo{number}{1} (\bibinfo{year}{2015}), \bibinfo{pages}{1--19}.
\newblock
\showISSN{21931127}


\bibitem[Truong et~al\mbox{.}(2020)]%
        {Truong2020}
\bibfield{author}{\bibinfo{person}{Charles Truong}, \bibinfo{person}{Laurent
  Oudre}, {and} \bibinfo{person}{Nicolas Vayatis}.}
  \bibinfo{year}{2020}\natexlab{}.
\newblock \showarticletitle{{Selective review of offline change point detection
  methods}}.
\newblock \bibinfo{journal}{\emph{Signal Processing}}  \bibinfo{volume}{167}
  (\bibinfo{year}{2020}).
\newblock
\showISSN{01651684}


\bibitem[Wei et~al\mbox{.}(2013)]%
        {Wei2013}
\bibfield{author}{\bibinfo{person}{Xinyu Wei}, \bibinfo{person}{Long Sha},
  \bibinfo{person}{Patrick Lucey}, \bibinfo{person}{Stuart Morgan}, {and}
  \bibinfo{person}{Sridha Sridharan}.} \bibinfo{year}{2013}\natexlab{}.
\newblock \showarticletitle{{Large-scale analysis of formations in soccer}}. In
  \bibinfo{booktitle}{\emph{International Conference on Digital Image
  Computing: Techniques and Applications}}.
\newblock
\showISBNx{9781479921263}


\bibitem[Wu et~al\mbox{.}(2019)]%
        {Wu2019}
\bibfield{author}{\bibinfo{person}{Yingcai Wu}, \bibinfo{person}{Xiao Xie},
  \bibinfo{person}{Jiachen Wang}, \bibinfo{person}{Dazhen Deng},
  \bibinfo{person}{Hongye Liang}, \bibinfo{person}{Hui Zhang},
  \bibinfo{person}{Shoubin Cheng}, {and} \bibinfo{person}{Wei Chen}.}
  \bibinfo{year}{2019}\natexlab{}.
\newblock \showarticletitle{{ForVizor: Visualizing spatio-temporal team
  formations in soccer}}.
\newblock \bibinfo{journal}{\emph{IEEE Transactions on Visualization and
  Computer Graphics}} \bibinfo{volume}{25}, \bibinfo{number}{1}
  (\bibinfo{year}{2019}), \bibinfo{pages}{65--75}.
\newblock
\showISSN{19410506}


\bibitem[Zhang and Chen(2017)]%
        {Zhang2017}
\bibfield{author}{\bibinfo{person}{Jingru Zhang} {and} \bibinfo{person}{Hao
  Chen}.} \bibinfo{year}{2017}\natexlab{}.
\newblock \showarticletitle{{Graph-based two-sample tests for data with
  repeated observations}}.
\newblock  (\bibinfo{year}{2017}).
\newblock
\showeprint[arxiv]{1711.04349}


\end{thebibliography}

\newpage
\appendix

\begin{table*}[hbt]
  \caption{Results of SoccerCPD with different CPD methods.}
  \label{tab:baselines}
  \begin{tabular}{ll|rr|rrrrrrr|rr}
    \toprule
    \multicolumn{2}{c|}{CPD methods} & \multicolumn{2}{c|}{\#. periods} & \multicolumn{7}{c|}{Proportion in playing time} & \multicolumn{2}{c}{Accuracy} \\
    FormCPD & RoleCPD & FP & RP & 3-4-3 & 3-5-2 & 4-4-2 & 4-2-3-1 & 4-3-3 & 4-1-3-2 & others & Formation & Position \\
    \midrule
    Avg. g-seg.     & Avg. g-seg.   & 864 & 2152 & 22.3\% & 12.2\% & 16.3\% & 20.9\% & 17.4\% & 5.8\% & 5.1\% & \textbf{72.4\%} & \textbf{86.5\%} \\
    Union g-seg.    & Avg. g-seg.   & 862 & 2263 & 20.5\% & 11.1\% & 13.2\% & 19.4\% & 20.2\% & 4.8\% & 10.9\% & 67.4\% & 85.4\% \\
    RBF kernel      & Avg. g-seg.   & 879 & 2278 & 21.6\% & 12.0\% & 16.6\% & 23.1\% & 15.9\% & 4.4\% & 6.4\% & 67.1\% & 83.8\% \\
    Cosine kernel   & Avg. g-seg.   & 883 & 2269 & 21.2\% & 12.0\% & 14.7\% & 19.7\% & 17.9\% & 8.0\% & 6.4\% & 67.1\% & 84.0\% \\
    Rank            & Avg. g-seg.   & 864 & 2223 & 17.8\% & 12.4\% & 15.3\% & 19.1\% & 22.5\% & 4.1\% & 8.7\% & 65.7\% & 85.1\% \\
    None            & 5-min seg.    & 807 & 8536 & 21.6\% & 12.4\% & 15.3\% & 23.4\% & 15.7\% & 6.3\% & 5.1\% & 59.1\% & 80.3\% \\
    \bottomrule
  \end{tabular}
\end{table*}

\section{Additional Explanation}

\subsection{Pipeline of the Framework} \label{pipeline}
\begin{figure}[!bth]
    \centering
    \includegraphics[width=0.48\textwidth]{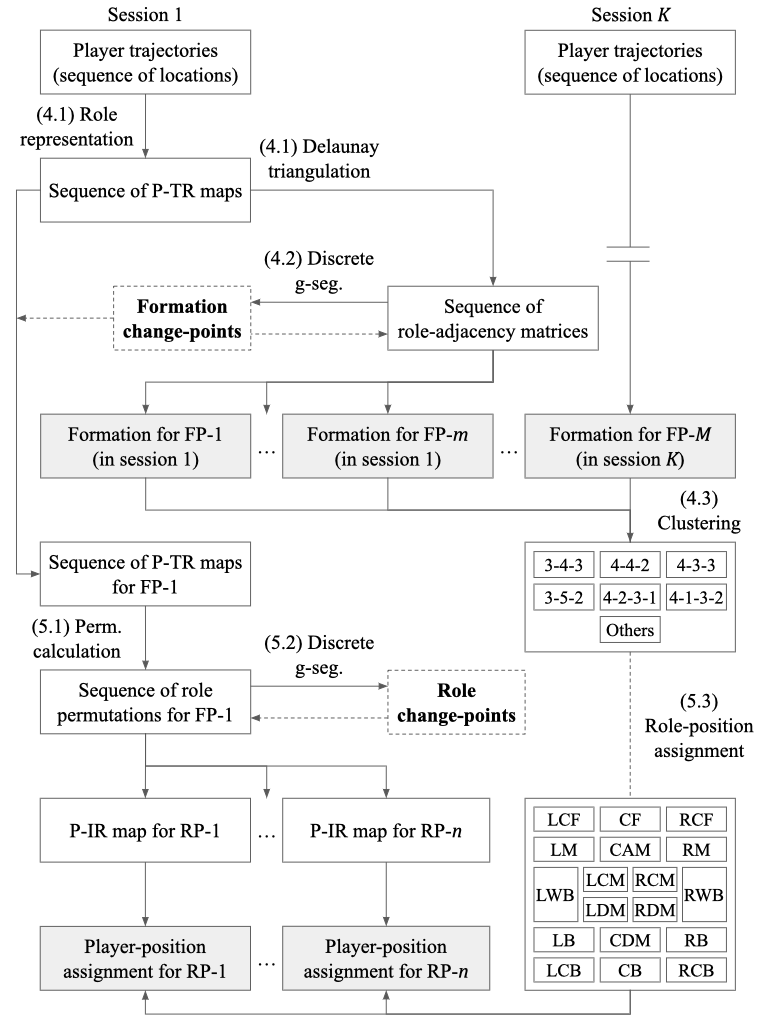}
    \caption{A block diagram for Sections \ref{formcpd} and \ref{rolecpd}. FP and RP are acronyms for formation and role periods, respectively.}
\label{fig:block_diagram}
\end{figure}

For better understanding of the whole pipeline, we attach a block diagram for Sections \ref{formcpd} and \ref{rolecpd} in Fig.~\ref{fig:block_diagram}. The upper part illustrates the process of applying FormCPD to a session (i.e., half of a match), and the lower part depicts the application of RoleCPD to a formation period resulting from the previous FormCPD. Note that formation clustering (Section~\ref{formcpd_cluster}) is the only step that uses data from multiple sessions. Also, formation and position labels on the lower right side of the diagram can vary depending on the dataset and the interpretation of the formation clusters.

\subsection{Details of Formation Clustering} \label{formcpd_cluster_details}
In Section~\ref{formcpd_cluster}, we proceed the formation clustering in two steps, since it is hard to distinguish 4-4-2 from 4-2-3-1 by a single step of clustering. First, we group 864 formations into 18 clusters where 18 is the minimum number of clusters that main formations except 4-4-2 and 4-2-3-1 are separated from one another. At the next step, the cluster in which 4-4-2 and 4-2-3-1 coexist (i.e., the mixture of Fig.~\ref{fig:formation_442} and Fig.~\ref{fig:formation_4231}) is further split into 18 sub-clusters, with one indicating 4-4-2 and all the others indicating 4-2-3-1. After merging clusters with less than 15 observations into ``others'' group and those indicating 4-2-3-1 into one group, we get seven formation groups as illustrated in Fig.~\ref{fig:clusters}.

\subsection{Python Implementation}
We have implemented the role representation algorithm using Python 3.8 on our own, while adopted the R package \texttt{gSeg}\footnote{https://rdrr.io/cran/gSeg/} developed by Chen et al.~\cite{Chen2015}, Chu et al.~\cite{Chu2019}, and Song et al.~\cite{Song2020} for discrete g-segmentation. To perform the whole process at once, we utilize the Python package \texttt{rpy2}\footnote{https://rpy2.github.io} to run the R script with \texttt{gSeg} inside our Python implementation. The process was executed on an Intel Core i7-8550U CPU with 16.0GB of memory. (Note that the M1 chip by Apple does not support binaries for \texttt{rpy2}'s API mode in Python, raising a memory error.) Our implementation is publicly accessible in a GitHub repository named \texttt{soccercpd}\footnote{https://github.com/pientist/soccercpd} with sample GPS data from an anonymized match.

\begin{figure*}[!htb]
    \centering
    \includegraphics[width=0.9\textwidth]{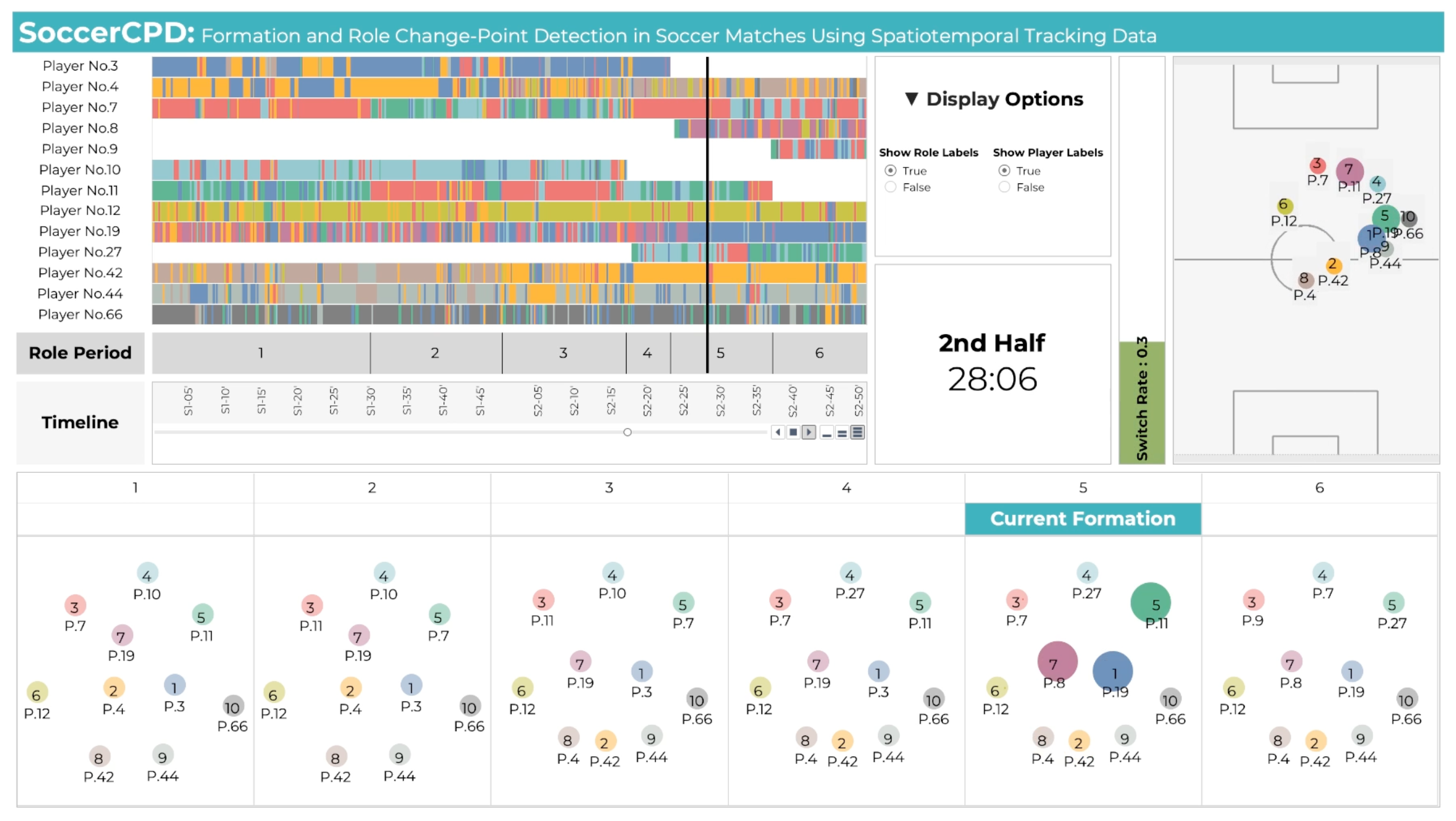}
    \caption{A snapshot of the Tableau animation of SoccerCPD.}
\label{fig:tableau_single}
\end{figure*}

\section{Comparison Against Baselines} \label{baseline_comparison}

To empirically justify our model selection, we have performed an additional experiment by comparing the results in Section~\ref{model_eval} against those using other baseline CPD methods.

Since FormCPD and RoleCPD handle different types of sequences, the sets of compatible baselines also differ from one another. FormCPD takes discrete high-dimensional data ($10 \times 10$ binary matrices) where parametric methods assuming a certain distribution on a Euclidean space show poor performances. Thus, we adopted nonparametric methods including kernel methods~\cite{Shawe-Taylor2004} and rank-based methods~\cite{Lung-Yut-Fong2015}. We also estimated a formation per session without finding a change-point as a naive baseline. For RoleCPD, the observations are permutations in a symmetric group where we cannot even define an inner product or an order. This means that kernel methods and rank-based methods are not applicable as well as the other off-the-shelf CPD methods for $\mathbb{R}^d$. Hence, we have only used the 5-minute smoothing to find instructed roles per time window as in Bialkowski et al.~\cite{Bialkowski2014-Large}.

To implement the CPD baselines, we utilize the Python package \texttt{ruptures}\footnote{https://centre-borelli.github.io/ruptures-docs/} developed by Truong et al.~\cite{Truong2020}. The applied methods are listed below.

\begin{itemize}
    \item \textbf{Avg. g-seg.} (for both): Discrete g-segmentation using the average scan statistic. (\texttt{gseg1\_discrete} in \texttt{gSeg})
    \item \textbf{Union g-seg.} (for both): Discrete g-segmentation using the union scan statistic (\texttt{gseg1\_discrete} in \texttt{gSeg}).
    \item \textbf{RBF kernel} (only for FormCPD): Kernel CPD with the radial basis function kernel (\texttt{CostRbf} in \texttt{ruptures}).
    \item \textbf{Cosine kernel} (only for FormCPD): Kernel CPD with the cosine similarity kernel  (\texttt{CostCosine} in \texttt{ruptures}).
    \item \textbf{Rank} (only for FormCPD): CPD using a rank transformation (\texttt{CostRank} in \texttt{ruptures}).
    \item \textbf{None} (only for FormCPD): Estimating a formation from the entire session without applying a CPD method.
    \item \textbf{5-min seg.} (only for RoleCPD): Segmenting a player period (obtained by splitting a session by player substitutions) into 5-minute windows and find the most frequent roles for each player per window.
\end{itemize}

Table~\ref{tab:baselines} shows the results of SoccerCPD for each method. All the baselines except for the non-CPD one split the total of 807 sessions into about 870 formation periods and 2,200 role periods. It clearly demonstrates that higher performances can be obtained when employing CPD methods instead of the naive segmentation. Moreover, discrete g-segmentation with the average scan statistic outperforms the other CPD methods.

\section{Tableau Visualization} \label{tableau}
Lastly, we visualize our results as an animation using Tableau 2020.4. Fig.~\ref{fig:tableau_single} is a snapshot of the generated animation. It dynamically shows all the following tactical information in one dashboard:
\begin{itemize}
    \item \textbf{Player trajectories} colored by temporary roles in the right.
    \item \textbf{Player-to-temporary-role assignments} as colored bar codes in the left timeline, divided into detected role periods.
    \item \textbf{Formation} per role period by the mean role locations in the bottom with the emphasis on the current formation.
    \item \textbf{Player-to-instructed-role assignments} per role period by annotating uniform numbers below the role circles.
    \item \textbf{Switching players} indicated by inflating the corresponding circles both in the formations and the pitch image.
\end{itemize}

We have published the full-version video on YouTube\footnote{https://www.youtube.com/watch?v=F-tvG1-MLps}. We expect that this visual interface effectively supports domain participants to review and analyze the match by tracking tactical changes and discovering switching patterns.

\end{document}